\documentclass[aps,prr,superscriptaddress, twocolumn]{revtex4-1}
\usepackage[english]{babel}
\usepackage[ansinew]{inputenc}
\usepackage{times}
\usepackage{graphicx}
\usepackage{graphics}
\usepackage{amsmath}
\usepackage{amsfonts}
\usepackage{amssymb}
\usepackage{dsfont}
\usepackage{epstopdf}
\usepackage{makeidx}
\usepackage{subfigure}
\usepackage{color}
\usepackage{pgf}
\usepackage{bm}
\usepackage{ulem}
\usepackage{tikz} 

\newcommand{\be}{\begin{equation}}
\newcommand{\ee}{\end{equation}}
\newcommand{\bea}{\begin{eqnarray}}
\newcommand{\eea}{\end{eqnarray}}

\newcommand\redsout{\bgroup\markoverwith{\textcolor{red}{\rule[0.5ex]{4pt}{0.8pt}}}\ULon}

\makeindex

\begin{document}

\title{Unconventional superconductivity as a synchronization problem in nuclear oscillator networks}

\author{V. Velasco} 

\affiliation{Instituto de Fisica, Universidade Federal do Rio de Janeiro, Caixa Postal
68528, Rio de Janeiro, Brazil}

\author{M. B. Silva Neto} 

\affiliation{Instituto de Fisica, Universidade Federal do Rio de Janeiro, Caixa Postal
68528, Rio de Janeiro, Brazil}

\begin{abstract}

We formulate the problem of unconventional $d-$wave superconductivity, 
with phase fluctuations, pseudogap phenomenon, and local Cooper pairs, 
in terms of a synchronization problem in random, quantum dissipative, 
elasto-nuclear oscillator networks. The nodes of the network correspond to 
{\it localized, collective quadrupolar vibrations} of nuclei-like, elastic inhomogeneities 
embedded in a dissipative medium. Electrons interacting with such vibrations 
form local Cooper pairs, with a superfluid $d-$wave pseudogap $\Delta_{PG}$, due 
to an effective, short range attractive interaction of $d_{x^2-y^2}$ character. 
Phase coherent, bulk superconductivity, with a $d-$wave gap $\Delta$, is 
stabilized when the oscillator network is asymptotically entangled in a nearly 
decoherence-free environment. Phase coherence will in turn be destroyed, 
at $T_c$, when the thermal noise becomes comparable to the coupling 
between oscillators, the superfluid density $K$. The $2\Delta/k_B T_c$ ratio 
is a function of Kuramoto's order parameter, $r=\sqrt{1-K_c/K}$, for the loss 
of synchronization at $K_c$, and is much larger than the nonuniversal $2\Delta_{PG}/k_B T^*$ 
ratio, where $T^*$ is the temperature at which $\Delta_{PG}$ is completely 
destroyed by thermal fluctuations. We discuss our findings in connection 
to the available data for various unconventionally high-temperature superconductors.

\end{abstract}

\maketitle

\section{Introduction}

Quadrupolar vibrations are ubiquitous in nature and often constitute the fundamental
normal modes of vibration in a plethora of different physical systems. They arise as 
the lowest frequency modes of vibration in wineglasses \cite{Jundt2006}, and also as 
the most relevant tectonic field for density perturbations induced by earthquake ruptures 
\cite{Harms2015}. Accelerating masses moving through spacetime produce ripples
that propagate as transverse, quadrupolar gravitational waves \cite{Einstein1918}, 
and quadrupolar vibrations of the inner crust in a neutron star controls its transient 
cooling, when coupled to a dissipative, outter thermal bath \cite{Inakura2019}. At 
smaller scales, quadrupolar surface vibrations in finite nuclei are known to contribute 
to the giant quadrupolar resonance \cite{Ligensa1966}. Most importantly, these very 
same quadrupolar surface vibrations contribute also to a remarkable emergent phenomenon 
in nuclear matter: superfluidity \cite{Nuclear-Superfluidity}. For finite nuclei, this is 
the mechanism behind the opening of superfluid gaps, $\Delta$, in nuclear spectra, 
for nucleons that minimize their energy, in the presence of a short range attractive 
nuclear potential, by moving in Cooper paired, time-reversed 
orbits \cite{Barranco1999}. 
 
The spontaneous emergence of collective behaviour in large oscillator networks is 
also a phenomenon that has applications in many branches of science. These include
the description of the synchronous flash of fireflies \cite{Ermentrout1991} and the 
generation of alpha rhythms in the brain \cite{Tass2003}. Huygen's pendulum clocks, 
weakly coupled through a wooden beam \cite{Bennett2002}, or metronomes 
sharing a common base \cite{Pantaleone2002} are examples of anti-phase and 
in-phase {\it classical} synchronization, respectively.
The dynamics of fast spins coupled to slow exchange interactions in XY spin 
glasses \cite{Jongen2001} and the frequency locking in superconducting Josephson 
junction arrays \cite{Wiesenfeld1998} are, on the other hand, examples of emergent 
collective behaviour in {\it quantum} oscillator networks. In all those cases, the model
that has become the simplest paradigm for the synchronization phenomenon is the
Kuramoto model \cite{Kuramoto1975}. It relies basically on two properties: i) the 
couplings between the node oscillators in the network: a superfluid density, $K$; and 
ii) the presence of white noise, quenched, $\delta$, or thermal, $k_B T$, provided by an 
environment. Then, partial or full synchronization is achieved when the couplings,
$K$, outgrow the noise, $\delta,k_B T$, while phase coherence is entirely lost otherwise. 
The synchronization phase transition in Kuramoto's model is described by a complex 
order parameter with a real part $r=0$, for no-, $0<r<1$, for partial-, and $r=1$, for 
full-synchronization \cite{Kuramoto1987}. 

%
\begin{figure}[!t]
	\begin{center}
		\includegraphics[scale = 0.39]{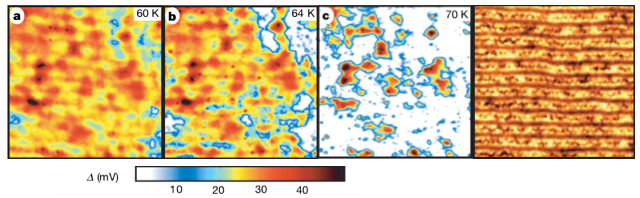}
	\end{center}
\caption{Adapted from ref. \cite{Yazdani2007}. Left: gap inhomogeneities 
deduced from STM spectra at: (a) $T=60$K (below), (b) $T=64$K (around), and (c) $T=70$K 
(above), the critical $T_c=65$K in overdoped Bi$_2$Sr$_2$CaCu$_2$O$_{8+\delta}$. 
Right: topographic STM image of the positions where the gaps were measured. The STM
images clearly show that local Cooper pairs form at random elastic inhomogeneities.}
\label{Fig-Inhomogeneities-BSCCO}
\end{figure}
%

In this work we formulate the problem of unconventional $d-$wave superconductivity
in terms of a synchronization problem in random, quantum dissipative, nuclear
oscillator networks. Our basic requirement is the existence of elastic 
inhomogeneities as the ones shown in Fig. \ref{Fig-Inhomogeneities-BSCCO} \cite{Yazdani2007}. 
We use elasticity theory to calculate the collective, normal modes of vibration 
at these elastic insertions and show that electrons couple to their low-lying 
quadrupolar mode. An effective, short range, $d_{x^2-y^2}-$wave attractive, 
two-particle interaction, leads to the formation of time-reversed, local 
Cooper pairs, in close analogy to nuclear superfluidity. We calculate the 
quantum wavefunctions for localized, collective quadrupolar vibrations forming a 
connected, nuclear oscillator network through their overlap, $K$, and show that, 
for $K$ stronger than the decoherence noise of the environment, $\delta,k_B T$, 
phase locking occurs and bulk superconductivity emerges. Finally, we calculate 
the bulk gap, $\Delta$, the transition temperature, $T_c$, and compare their ratio, 
$2\Delta/k_B T_c$, given as a function of Kuramoto's order parameter, $r$, 
to the available data for various compounds. 

\section{Quadrupolar Vibrations}

We begin by showing that, already at the classical level, the lowest energetic mode of 
vibration for an elastic sphere embedded in a infinite homogeneous medium, with different
elasto-mechanical properties such as the Lam\` e parameters, $\lambda$ and $\mu$, and 
the density, $\rho$, is the quadrupole mode. For that we need to solve Navier's equation for the 
displacement
\begin{equation}
{\bf u}(r,\theta,\phi,t) = {\bf u}(r,\theta,\phi)\exp(i\Omega t), 
\label{Classical-Displacement}
\end{equation}
which, for the simplest case of a uniform, elastic, isotropic, ideal medium, reads \cite{Landau-Lifshitz}
\begin{equation}
    \rho\frac{\partial^2{\bf u}}{\partial t^2}+\mu\nabla\times(\nabla\times{\bf u})-
    (\lambda+2\mu)\nabla(\nabla\cdot{\bf u})={\bf F}.
    \label{Navier}
\end{equation}
In eq. (\ref{Navier}) ${\bf F}$ represents all volume and surface equilibrium external forces
that are provided by the crystalline host enclosing the spherical elastic insertion. The stationary, 
homogeneous, ${\bf F}=0$, solutions to (\ref{Navier}) are of the form 
\begin{equation*}
{\bf u}(r,\theta,\phi) = u_r\hat{\bf r} + u_{\theta}\hat{\mathbf{\theta}} + u_{\phi}\hat{\mathbf{\phi}},
\end{equation*} 
and two types of vibrations arise from the continuity conditions: torsional and spheroidal. 
Torsional vibrations are characterized by the vanishing of the radial displacement, $u_r=0$, 
and by the vanishing of divergent displacements, $\nabla\cdot{\bf u}=0$. Spheroidal
vibrations, on the other hand, are characterized by the vanishing of the radial part of 
circulation displacements, $\nabla \times {\bf u}$. In what follows we shall be interested 
in considering radial spheroidal solutions to eq. (\ref{Navier}) of the kind \cite{Dubroviskiy1981}
\begin{eqnarray*}
    u_r &=& \sum_{\ell,m}\left[\frac{A_{\ell,m}d_1(k_pr)}{k_pr} + \frac{B_{\ell,m}\ell(\ell+1)b_\ell(k_sr)}{k_sr}\right]Y_\ell^m, \nonumber \\
    u_{\theta} &=& \sum_{\ell,m}\left[\frac{A_{\ell,m}b_\ell(k_pr)}{k_pr} + \frac{B_{\ell,m}d_2(k_sr)}{k_sr}\right] 
    \frac{\partial Y_\ell^m}{\partial \theta},\nonumber \\ 
    u_{\phi} &=& \sum_{\ell,m}\left[\frac{A_{\ell,m}b_\ell(k_pr)}{k_pr}+ \frac{B_{\ell,m}d_2(k_sr)}{k_sr}\right]
    \frac{1}{\sin{(\theta)}}\frac{\partial Y_\ell^m}{\partial \phi},\nonumber
\end{eqnarray*}
where $k_s^2 = \Omega^2/c_s^2$ and $k_p^2 = \Omega^2/c_p^2$, with $c_s$ and $c_p$ being 
the velocities of the transverse and longitudinal elastic waves, respectively \cite{Dubroviskiy1981}. 
The coefficients $A_{\ell,m}$ and $B_{\ell,m}$ are fixed by the boundary conditions of continuity for 
the displacement, ${\bf u}$, and by the radial, $\sigma_{rr}$, and shearing, $\tau_{r\theta}$ and 
$\tau_{r\phi}$, stresses, at the radius of the inhomogeneous insertion, $r=R_0$, relevant when 
${\bf F}\neq 0$. As usual, $Y_\ell^m(\theta, \phi)$ are the spherical harmonics, and we have
defined the function $b_\ell(z) = h_{\ell}(z)$, for vibrations of the crystalling host, outside the 
sphere, and the function $b_\ell(z) = j_{\ell}(z)$ for vibrations of the embedded elastic inhomogeneity, with 
$j_\ell(z)$ and $h_\ell(z)$ being the spherical Bessel and Hankel functions, respectively. The 
use of spherical Bessel and Hankel functions ensures that the displacement, ${\bf u}$, vanishes 
both at the origin and at infinity, which are natural boundary conditions for any finite elastic 
deformation. Finally, we have also defined $d_1(z) = \ell b_\ell(z) - zb_{\ell+1}(z)$, 
and $d_2(z) = (\ell+1)b_\ell(z) - zb_{\ell+1}(z)$, for compactness.

%
\begin{figure}[!t]
\begin{center}
\includegraphics[scale = 0.37]{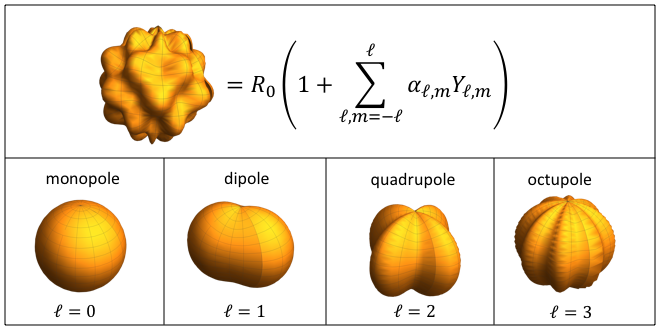}
\end{center}
\caption{Top: the general shape of any elastic inhomogeneity can be described through collective 
coordinates, $\alpha_{\ell,m}$, as a multipolar expansion in terms of spherical harmonics, $Y_{\ell,m}$. 
Bottom: The first terms of the expansion showing: monopole ($\ell=0$), dipole ($\ell=1$), 
quadrupole ($\ell=2$), and octupole ($\ell=3$) terms. The monopole $\ell=0$ correspond to
volume changes while the dipole $\ell=1$ to translations of the elastic insertion, being, as
such, too energetically costly. The lowest normal mode is thus the quadrupole $\ell=2$.}
\label{Fig-Collective-Coordinates}
\end{figure}
%

The normal frequencies, $\Omega$, are found, as usual, from the zeroes of the determinant for 
the system of coupled equations defined by eq. (\ref{Navier}), and their real or complex characters
are determined by the ratio between the shear moduli of the enclosing medium and the inclusion,
$p=\beta^2\eta$ \cite{Dubroviskiy1981}. Here $\eta=\rho_e/\rho_i$ is the ratio between the external 
($e$) and internal ($i$) densities, and $\beta= c_{se}/c_{si}$ is the ratio between the transverse 
elastic velocities outside and inside the sphere. For the case of freely vibrating elastic objects (or 
for a very soft enclosing medium) then $p\rightarrow 0$ and all normal frequencies are real, $\Omega\in{\cal R}$:
once excited the insertion is set into permanent, stationary vibration. For radial-spheroidal elastic 
vibrations embedded in an equally ellastic, homogeneous crystalline host, however, then $p\approx 1$ 
and we end up with the transcendental equation \cite{Dubroviskiy1981}
\begin{eqnarray}
    \ell(\eta - 1) + g_{e}(\zeta \chi) - \eta g_{i}(\alpha \zeta \chi) = 0,
    \label{Normal-Frequencies}
\end{eqnarray}
where $\chi = \Omega R_0/c_{se}$ is a dimensionless frequency written in terms of the transverse 
elastic velocity in the exterior, $c_{se}$, while $\alpha= c_{pe}/c_{pi}$ and $\zeta= c_{se}/c_{pe}$ 
are ratios between the transverse and longitudinal elastic velocities outside and inside the sphere. 
The functions $g_i(z)=z j_{\ell+1}(z)/j_\ell(z)$, and $g_e(z)=z h_{\ell+1}(z)/h_\ell(z)$, are also defined 
inside and outside the sphere, respectively \cite{Dubroviskiy1981}. In this case, eq. (\ref{Normal-Frequencies}) produces,
instead, complex solutions, $\Omega\in{\cal C}$, with a real part, ${\cal R}e[\Omega]\neq 0$, that
sets the natural frequency of vibration and an imaginary part, ${\cal I}m[\Omega]=\gamma\neq 0$, that provides
damping, which can be naturally understood as the decay of the localized, normal vibrations into 
divergent spherical elastic waves of the enclosing host \cite{Dubroviskiy1981}. Finally, the lowest ${\cal R}e[\Omega]\neq 0$ 
mode of vibration corresponds to $\ell=2$, {\it localized, collective quadrupolar vibrations}, see bottom 
of Fig. \ref{Fig-Collective-Coordinates}, just like for the vibrations of wineglasses \cite{Jundt2006}, or at the 
core of neutron stars \cite{Inakura2019}, or at the surface of nuclei \cite{Barranco1999}. The monopole, 
$\ell=0$, is a breathing mode associated to volume changes at the insertion, see 
bottom of Fig. \ref{Fig-Collective-Coordinates}, while dipole, $\ell=1$, corresponds to translations of 
the center of mass of the insertion, see bottom of Fig. \ref{Fig-Collective-Coordinates}, and both 
$\ell=0,1$ solutions are thus too energetically costly. 

\section{The $d_{x^2-y^2}$ Particle Vibration Coupling}

Once we established that the lowest mode of vibration is the $\ell=2$ 
quadrupole mode, let us now proceed and narrow the true vibrational ground state 
down to the $d_{x^2-y^2}$ mode, which will be always valid for layered, anisotropic 
systems such as the high temperature cuprates. For that, we recall that 
quadrupolar deformations from a spherical equilibrium can be parametrized in terms 
of collective coordinates $\alpha_{2,m}$ \cite{Nuclear-Structure}
\begin{equation}
    R=R_0\left(1+\sum_{m=-2}^{m=+2}\alpha_{2,m}Y_{2}^{m}\right).
\end{equation}
The equilibrium configuration requires that $\alpha_{2,m}=\alpha_{2,-m}$ and, since the
radius of a sphere is always a real quantity, $R\in{\cal R}$, one must also impose that
$\alpha^*_{2,m}=(-)^m\alpha_{2,-m}$. These constraints lead to $\alpha_{2,1}=\alpha_{2,-1}=0$,
eliminating deformations associated to combinations of $Y_{2}^{+1}$ and $Y_{2}^{-1}$. 
Furthermore, the $R\in{\cal R}$ constraint also leads to $\alpha_{2,0}\in{\cal R}$ and 
$\alpha_{2,2}=\alpha_{2,-2}\in{\cal R}$, eliminating also deformations associated 
to combinations of $Y_{2}^{+2}$ and $Y_{2}^{-2}$ containing the imaginary unit, $i$. 
Altogether, these constraints exclude completely any deformations associated to the 
three $t_{2g}$ orbitals
\begin{eqnarray}
Y_{2,1c}&=&\frac{1}{\sqrt{2}}(Y_2^{-1} - Y_{2}^{1})=\sqrt{\frac{15}{4\pi}}(\frac{xz}{r^2}),\nonumber\\ 
Y_{2,1s}&=&\frac{i}{\sqrt{2}}(Y_2^{-1} + Y_{2}^{1})=\sqrt{\frac{15}{4\pi}}(\frac{yz}{r^2}),\nonumber\\
Y_{2,2s}&=&\frac{i}{\sqrt{2}}(Y_2^{-2} - Y_{2}^{2})=\sqrt{\frac{15}{4\pi}}(\frac{xy}{r^2}).\nonumber
\label{t2g-Spherical-Harmonics}
\end{eqnarray}
The $\alpha_{2,0}\in{\cal R}$ and $\alpha_{2,2}=\alpha_{2,-2}\in{\cal R}$ constraints leave
us then with only two symmetry allowed, real, $e_g$ orbitals 
\begin{eqnarray}
    Y_{2,0} &=& Y_{2}^0 = \sqrt{\frac{5}{16\pi}}(3\frac{z^2}{r^2} - 1),\nonumber\\
    Y_{2,2c} &=& \frac{1}{\sqrt{2}}(Y_2^{-2} + Y_{2}^{2}) = \sqrt{\frac{15}{16\pi}}(\frac{x^2 - y^2}{r^2}),
\label{eg-Spherical-Harmonics}
\end{eqnarray}
where $s,c$ stand for sine and cosine, respectively. 

\begin{figure}[!t]
\begin{center}
\includegraphics[scale = 0.29]{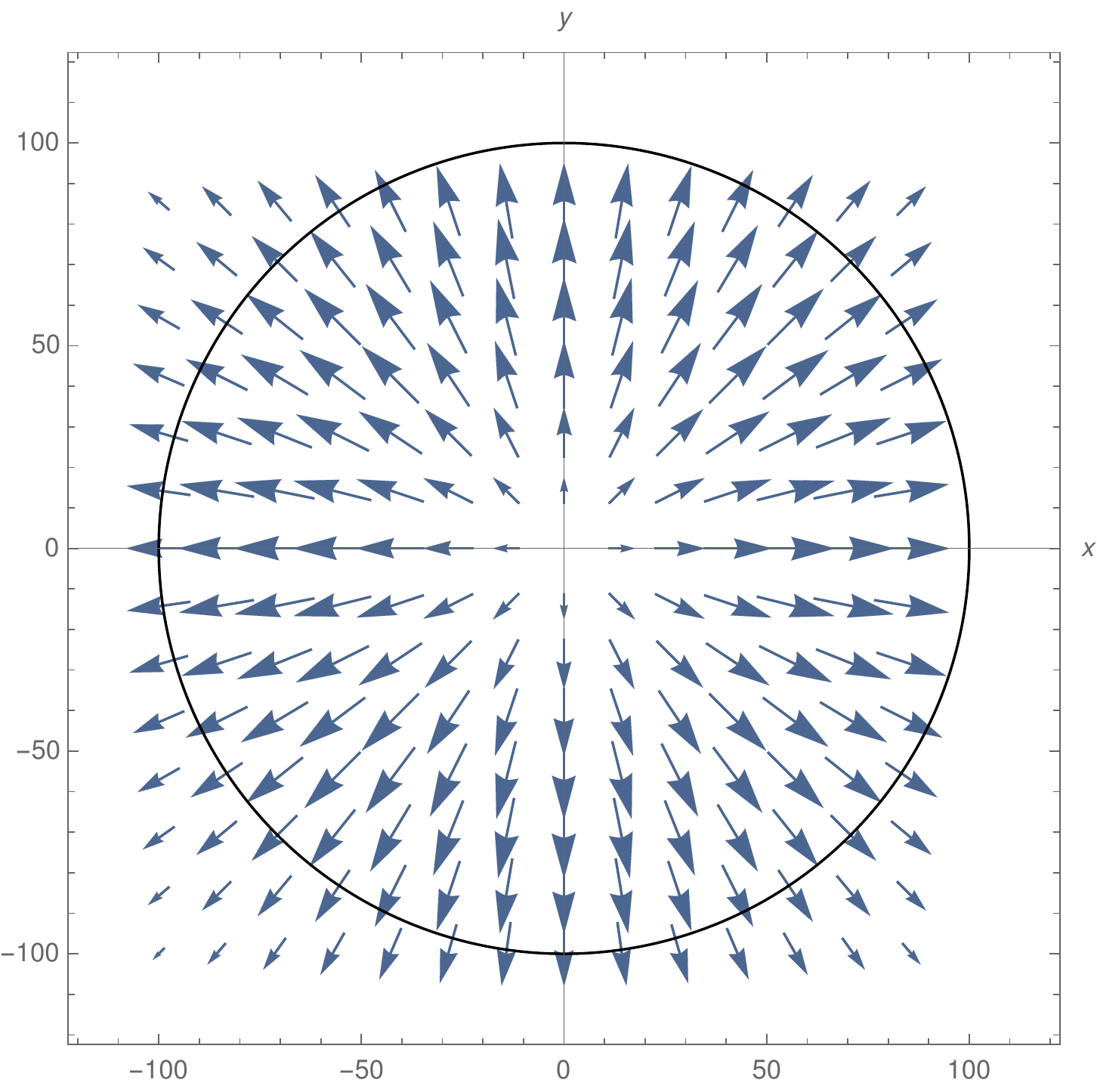}
\includegraphics[scale = 0.26]{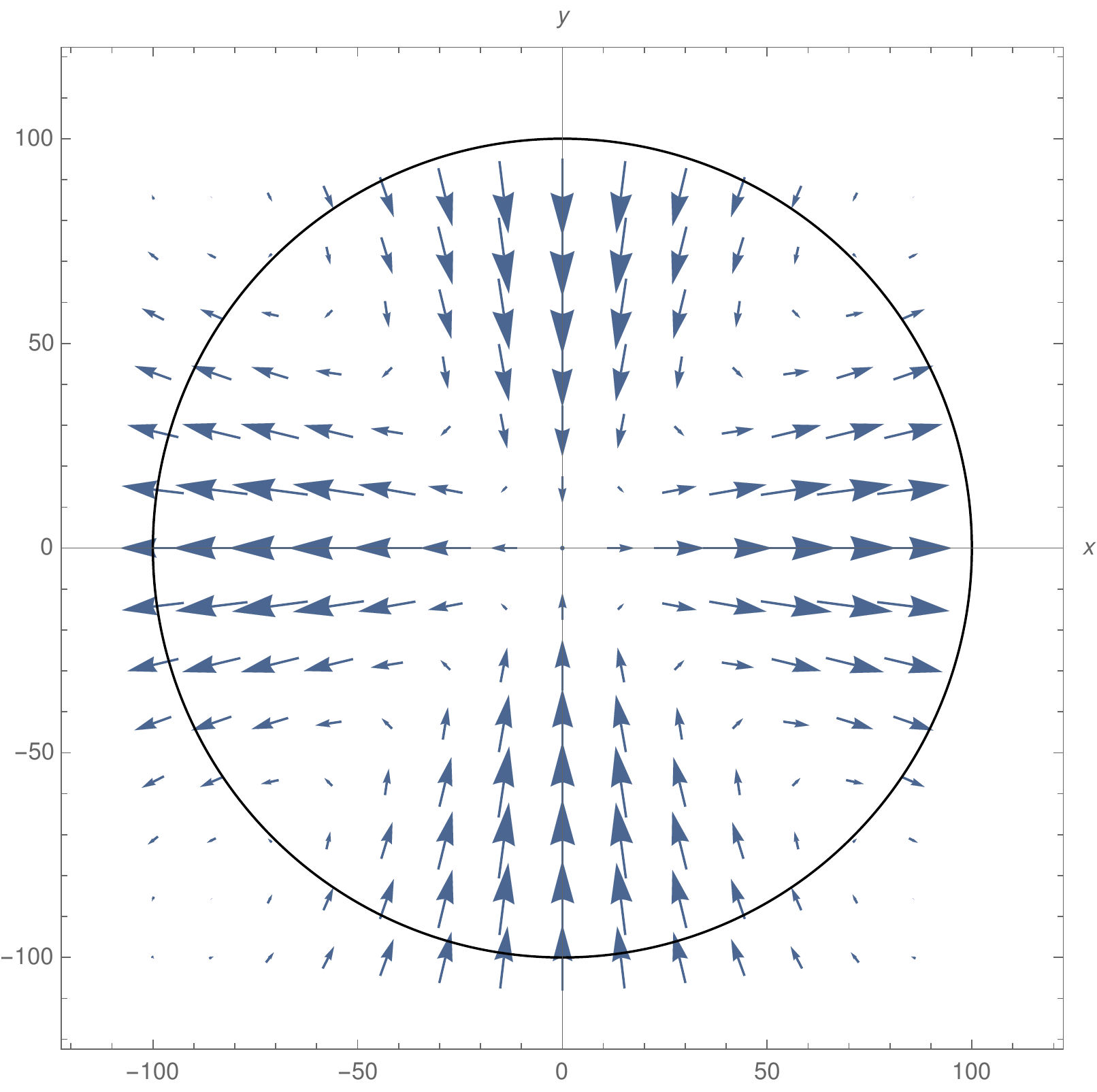}
\end{center}
\caption{Equatorial cuts for the $Y_{2,0}$ (left) and $Y_{2,2c}$ (right) radial-like
spheroidal displacements of eq. (\ref{eg-Spherical-Harmonics}). For layered, anisotropic
systems, such as the cuprates, the configuration which minimizes the elastic
energy is $Y_{2,2c}$ (right), and the true ground state for elasto-nuclear quadrupolar 
vibrations corresponds to the $d_{x^2-y^2}$ symmetry.}
\label{Fig-Displacements}
\end{figure}

For isotropic or weakly anisotropic systems the two $e_g$ deformations in 
eq. (\ref{eg-Spherical-Harmonics}) are degenerate (or nearly) and the lowest 
elastic quadrupolar vibrations should contain admixtures between the two  
$Y_{2,0}$ and $Y_{2,2c}$ orbitals. This might, perhaps, be relevant to 
the description of the two-gap structure observed by ARPES in some cuprates 
\cite{Yoshida2009}. For strongly anisotropic, weakly-coupled, layered structured 
systems, however, such as the majority of the high-temperature cuprates, the degeneracy 
between the two $e_g$ orbitals is lifted because of the large energy cost for 
planar stretches required by the $Y_{2,0}$ oblate and prolate deformations 
(see Fig. \ref{Fig-Displacements}, left). In this case, the true, lowest order, 
radial-spheroidal vibration mode is the quadrupolar $d_{x^2-y^2}-$mode, 
associated to the $Y_{2,2c}$ orbital (see Fig. \ref{Fig-Displacements}, right). 
It is worth point out that such elastic $d_{x^2-y^2}$ mode breaks the $C_4$ 
rotational symmetry of the lattice and may be related to the nematicity recently 
observed in several layered cuprates, in the pseudogap phase \cite{Daou2010,Fradkin2010,Sato2017}. 

%
%

Before we are able to write down the particle vibration coupling (PVC) Hamiltonian
we must first quantize quadrupolar vibrations. A quantum mechanical Hamiltonian for 
deformations $Y_{2,0}$ and $Y_{2,2c}$ can be written in terms of the 
Bohr shape variables, $\alpha_{2,0}=\beta\cos\gamma$ and 
$\alpha_{2,2}=\alpha_{2,-2}=(\beta/\sqrt{2})\sin\gamma$ \cite{Nuclear-Structure}, 
such that $\beta^2=\sum_{m=-2}^{m=2}\alpha^*_{2,m}\alpha_{2,m}$ and thus
\begin{equation}
    H=-\frac{\hbar^2}{2B}\sum_m \frac{\partial^2}{\partial\alpha_{2,m}\partial\alpha^*_{2,m}}+
    \frac{C}{2}\beta^2,
\end{equation}
where $B$ is some inertial parameter for the quadrupolar vibration of normal frequency 
$\Omega=\sqrt{C/B}$. This is a five dimensional problem in which the separable 
eigenfunctions of $H$, $\Psi(\beta,\gamma,\underline{\theta})=f(\beta)\Phi(\gamma,\underline{\theta})$, 
are written not only in terms of the two Bohr shape variables, $\beta$ and $\gamma$, but also 
in terms of three Euler angles, $\theta_{i=1\dots 3}$. In what follows we fix the crystal-to-laboratory 
reference frames, through a convenient choice for the Euler angles, $\underline{\theta}$, and we find for the 
radial part 
\begin{equation}
f(\beta)=F_{n,\tau}\beta^\tau \; e^{-s^2\beta^2/2}L_n^{\tau+3/2}(s^2\beta^2),
\label{Radial-Wave-Function}
\end{equation}
with spectrum $E_{n,\tau}=\hbar\Omega(N+5/2)$, and $N=2n+\tau$ \cite{Corrigan1976}. 
Here $F_{n,\tau}$ is a normalization constant, $n$ is the index of the radial solution, 
$s=(BC/\hbar^2)^{1/4}$ is the oscillator stiffness, $L_n^{\tau+3/2}$ is a Laguerre polynomial, 
and $\tau$ is the seniority \cite{Corrigan1976}.  

Particles of effective mass, $m^*$, moving in the vicinity of an elastic inhomogeneity are 
subject to a local potential, $U(r)$, entering the single-particle Schr\" odinger's equation as 
\begin{equation}
\left[-\frac{\hbar^2\nabla^2}{2m^*}+U(r)\right]\varphi_{{\bf k}}({\bf r})=\epsilon({{\bf k}})\varphi_{{\bf k}}({\bf r}),
\label{Schroedinger}
\end{equation}
from which one obtains the single-particle wave functions, $\varphi_{{\bf k}}({\bf r})$, and
single-particle energies, $\epsilon({\bf k})$. The linear deformation potential for small 
displacements, $\alpha^2\ll\alpha$, reads  \cite{Nuclear-Superfluidity}
\begin{equation*}
    \delta U(r)=-R_0\frac{\partial U(r)}{\partial r}\sum_{m}\alpha_{2,m}Y_{2,m},
\end{equation*}
and the matrix element for a process where an electron, at initial state ${\bf k}$, 
scatters off an elastic insertion, setting it into vibration of the $Y_{2,2c}$ 
type, and goes into a final state ${\bf k}^\prime$ is  \cite{Nuclear-Superfluidity}
\begin{equation}
    \Gamma_{{\bf k},{\bf k}^\prime}=-\Gamma_0
    \int d^3{\bf r}\, \varphi_{{\bf k}^\prime}^*({\bf r}) \left[R_0\frac{\partial U(r)}{\partial r}Y_{2,2c}\right]
    \varphi_{{\bf k}}({\bf r}),
    \label{PVC-Matrix-Element}
\end{equation}
where $\varphi_{{\bf k}}({\bf r}),\varphi_{{\bf k}^\prime}^*({\bf r})$ are the single-particle wave functions 
solutions to eq. (\ref{Schroedinger}). Here
\begin{equation*}
\Gamma_0=\sqrt{\frac{\hbar}{2B\Omega}}\equiv\langle 2,2c|\hat{\alpha}_{2,2c}|0,0\rangle,
\end{equation*}
is the reduced matrix element for the quadrupolar deformations, $\hat{\alpha}_{2,2c}$, 
which, in second quantized form and in the Heisenberg picture, are given as
\begin{equation}
\hat{\alpha}_{2,2c}(t)=\sqrt{\frac{\hbar}{2B\Omega}}(b^\dagger_{2,2c}e^{-i\Omega t}+b_{2,\overline{2c}}e^{i\Omega t}),
\label{Secon-Quantization-Displacements}
\end{equation}
where $b^\dagger_{2,2c}$ and $b_{2,2c}$ are bosonic, phonon creation and anihilation 
operators for the $Y_{2,2c}$ quadrupolar vibrational modes with frequencies 
$\Omega=\sqrt{C/B}$. We recognize eq. (\ref{Secon-Quantization-Displacements})
as the second quantized, quantum mechanical version of eq. (\ref{Classical-Displacement}).

Within the adiabatic, Born-Oppenheimer approximation the particle motion and quadrupolar
vibrations occur at very different time scales and to all purposes the phases $e^{\pm i\Omega t}$ 
appearing in eq. (\ref{Secon-Quantization-Displacements}) can be omitted. We are then ready 
to write down the {\it local} PVC Hamiltonian between conduction electrons and the $d_{x^2-y^2}$ localized, 
collective quadrupolar vibrations in Born-Oppenheimer approximation as
\begin{equation}
H_{pvc}=\sum_{{\bf k},{\bf k}^\prime,\sigma}\Gamma_{{\bf k},{\bf k}^\prime}
c^\dagger_{{\bf k}^\prime,\sigma}c_{{\bf k},\sigma}(b^\dagger_{2,2c}+ b_{2,\overline{2c}}),
\label{PVC}
\end{equation}
where $c^\dagger_{{\bf k},\sigma},c_{{\bf k},\sigma}$ are the usual fermionic, creation and
anihilation operators for electrons with wavevector ${\bf k}$ and spin $\sigma$, 
associated to the single-particle wavefunctions $\varphi_{\bf k}(r)$, and with dispersion 
given by the single-particle energies $\epsilon({\bf k})$.

\section{Local Quadrupolar Superconductivity}

Consider now the problem of a band of conduction electrons interacting with an isolated, localized, 
collective quadrupolar vibration, as described by the total Hamiltonian
\begin{equation*}
H=\sum_{{\bf k},\sigma}\xi_{\bf k}c^\dagger_{{\bf k},\sigma}c_{{\bf k},\sigma}+
\hbar\Omega\left(b^\dagger_{2,2c}b_{2,2c}+\frac{5}{2}\right)+H_{pvc},
\end{equation*}
where $\langle b^\dagger_{2,2c}b_{2,2c}\rangle=N$ and $\xi_{\bf k}=\epsilon({\bf k})-\mu$, 
with dispersion relation $\epsilon({\bf k})$ relative to the chemical potential $\mu$. The 
particle-vibration-coupling in (\ref{PVC}) produces, in second order perturbation theory, an 
effective two-particle interaction
\begin{equation*}
    V_{{\bf k},{\bf k}^\prime}=|\Gamma_{{\bf k},{\bf k}^\prime}|^2
     \frac{\hbar\Omega}{(\xi_{\bf k}-\xi_{{\bf k}^\prime})^2-(\hbar\Omega)^2},
\end{equation*}
that is attractive for electrons close to the Fermi surface, 
$|\xi_{\bf k}|,|\xi_{{\bf k}^\prime}|\ll\hbar\Omega$. The explicit form of such two-particle 
interaction is obtained through a canonical transformation to eliminate the phonons 
producing a BCS Hamiltonian \cite{Bardeen1957}
\begin{equation*}
    H=\sum_{{\bf k},\sigma}\xi_{\bf k}c^\dagger_{{\bf k},\sigma}c_{{\bf k},\sigma}+
    \sum_{{\bf k},{\bf k}^\prime}V_{{\bf k},{\bf k}^\prime}
    c^\dagger_{{\bf k},\uparrow}c^\dagger_{-{\bf k},\downarrow}c_{-{\bf k}^\prime,\downarrow}c_{{\bf k}^\prime,\uparrow},
\end{equation*}
showing that electrons minimize their energy by moving in Cooper paired, time-reversed orbits. 
From this BCS Hamiltonian one makes the usual mean-field decoupling of the quartic interaction 
leading to the self-consistent gap equation
\begin{equation}
    \Delta_{{\bf k}}=-\sum_{{\bf k}^\prime}
    \frac{V_{{\bf k},{\bf k}^\prime}\Delta_{{\bf k}^\prime}}{2\sqrt{\xi^2_{{\bf k}^\prime}+|\Delta_{{\bf k^\prime}}|^2}}
    \tanh{\left(\frac{\sqrt{\xi^2_{{\bf k}^\prime}+|\Delta_{{\bf k^\prime}}|^2}}{2k_B T}\right)}.
\label{Gap-Equation}
\end{equation}
Since by construction $V_{{\bf k},{\bf k}^\prime}$ is separable we may write \cite{Gogny1975,Kennedy1964,Sedrakian2013}
\begin{eqnarray}
V_{{\bf k},{\bf k}^\prime}&=&-V_0\eta(\hat{\bf k})\eta(\hat{\bf k}^\prime)w(k)w(k^\prime)
\Theta(|\xi_{\bf k}|-\hbar\Omega)\Theta(|\xi_{{\bf k}^\prime}|-\hbar\Omega),\nonumber\\
\Delta_{{\bf k}}&=&\Delta_{PG}\,\eta(\hat{\bf k})w(k),\nonumber
\end{eqnarray}
where $\eta(\hat{\bf k})=\cos{k_x}-\cos{k_y}$ is the $d_{x^2-y^2}$ anisotropy form factor 
associated to the $Y_{2,2c}$ deformation and
\begin{equation}
w(k)=(b\sqrt{\pi})^3 \, e^{-b^2 k^2/2}
\end{equation}
is a momentum dependent form factor 
\cite{Sedrakian2013,Chappert2015}, obtained from the Fourier transform of a Gogny-type 
short range interaction \cite{Gogny1975}, in terms of a material dependent parameter 
$b\sim 1/R_0$. At zero temperature, $T=0$, the pseudogap equation reads 
\begin{equation*}
    1=V_0\sum_{{\bf k}^\prime}
    \frac{\eta^2(\hat{\bf k}^\prime)w^2(k^\prime)\Theta(|\xi_{{\bf k}^\prime}|-\hbar\Omega)}
    {2\sqrt{\xi^2_{{\bf k}^\prime}+\eta^2(\hat{\bf k}^\prime)w^2(k^\prime)\Delta_{PG}^2}}.
\end{equation*}
As usual we perform the average $\eta^2(\hat{\bf k}^\prime)\rightarrow\langle\eta^2\rangle_{FS}$, 
where $\langle\cdots\rangle_{FS}$ stands for angular average over the Fermi surface, and since 
$\hbar\Omega\ll \epsilon_F$ we can set $w^2(k)\rightarrow w^2(k_F)$ to arrive at 
the local (pseudo) gap at $T=0$
\begin{equation*}
\Delta_{PG} = \frac{2\hbar\Omega}{\sqrt{\left\langle \eta^2 \right\rangle_{FS}} w(k_F)}
\exp{\left\{-\frac{1}{\lambda\left\langle \eta^2 \right\rangle_{FS}w^2(k_F)}\right\}},
\end{equation*}
where $\lambda = N(\epsilon_F)V_0$ is proportional to the density of states at the 
Fermi level. At the temperature $T^*$ the local pseudogap, $\Delta_{PG}$, is 
completely destroyed by thermal fluctuations. In order to calculate $T^*$ one 
needs to solve the gap equation (\ref{Gap-Equation}) for $\Delta_{PG}(T^*)=0$ 
\begin{equation*}
    1=V_0\sum_{{\bf k}^\prime}
    \frac{\eta^2(\hat{\bf k}^\prime)w^2(k^\prime)
    \tanh{\left(|\xi_{{\bf k}^\prime}|/2k_B T^*\right)}\Theta(|\xi_{{\bf k}^\prime}|-\hbar\Omega)}
    {2|\xi_{{\bf k}^\prime}|},
\end{equation*}
whose solution is, after performing the same averaging procedures as done for $\Delta_{PG}$, given by
\begin{equation}
k_B T^*=1.13\,\hbar\Omega\,\exp{\left\{-\frac{1}{\lambda\left\langle \eta^2 \right\rangle_{FS}w^2(k_F)}\right\}}.
\end{equation}
We are now ready to calculate the pseudo-gap-to-$T^*$ ratio 
\begin{equation}
\frac{2\Delta_{PG}}{k_B T^*}=\frac{3.53}{\sqrt{\left\langle \eta^2 \right\rangle_{FS}} w(k_F)},
\label{Pseudogap-to-T-ratio}
\end{equation}
which is a nonuniversal ratio, due to the factor $w(k_F)$, larger than 
BCS's $3.53$, and can, in fact, be as large as $8$ \cite{Yazdani2007}.

%
\begin{figure}[!t]
\begin{center}
\includegraphics[scale = 0.26]{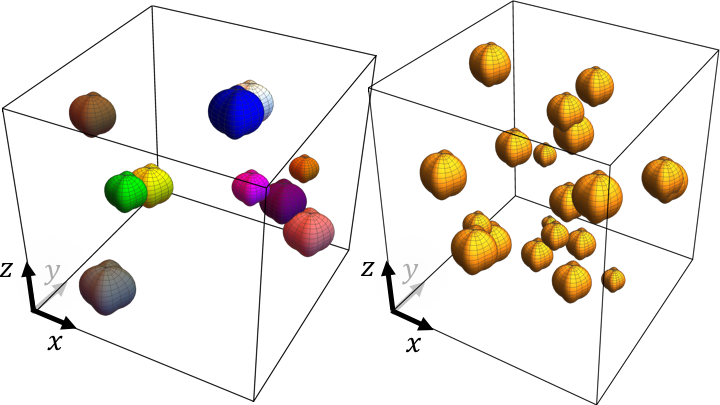}
\end{center}
\caption{(Color online) $-$ The quadrupolar oscillator network $-$ a random set of elasto-nuclear, quadrupolar 
inhomogeneities of arbitrary radii, $R_0^i$, and normal frequencies, $\Omega_i$. (Left) For 
$K<K_c$, dilute, distant insertions, $r=0$ and all phases (different colors) are independent. 
(Right) For $K>K_c$, insertions are more abundant and closer together, $r\rightarrow 1$ and 
phase-locking occurs (same color).}
\label{Fig-Quadrupolar-Network}
\end{figure}
%

\section{The Quadrupolar Oscillator Network}

Consider now a large number, $N$, of mutually interacting quadrupolar oscillators with distributed 
natural frequencies as shown in Fig. \ref{Fig-Quadrupolar-Network}. We want to establish the 
conditions for the spontaneous emergence of synchronization, or phase locking, in our 
quadrupolar oscillator network, as well as its consequences to unconventional, bulk 
superconductivity. For that  purpose, we write down the coupled Hamiltonian  
\begin{equation}
    H=-\frac{1}{2B}\hat{\Pi}_\alpha^\dagger\hat{\Pi}_\alpha+\frac{B}{2}\hat{\alpha}^\dagger\cdot{\cal H}\cdot\hat{\alpha},
    \label{Oscillator-Network}
\end{equation}
where $\hat{\Pi}_\alpha=-i\hbar\partial/\partial\hat{\alpha}$
and ${\cal H}_{ij}=\Omega_i^2\delta_{ij}+K_{ij}(1-\delta_{ij})$ \cite{Zambrini2013}. 
Here $\Omega_i$ are the quadrupolar natural frequencies, distributed according 
to a Lorentzian spectral distribution, 
\begin{equation}
g(w)=\frac{1}{\pi}\frac{\delta}{(w-\Omega)^2+\delta^2}, 
\label{Spectral-Density}
\end{equation}
that is symmetric with respect to $\Omega$, with a quenched spread, $\delta$, 
determining the amount of disorder in the material, while 
\begin{equation}
K_{ij}=\int d^3{\bf r}\; \Psi_{{\bf r}_i}^*(\beta,\gamma,\underline{\theta})
\Psi_{{\bf r}_j}(\beta,\gamma,\underline{\theta})\sim e^{-s^2|{\bf r}_i-{\bf r}_j|^2/2},
\label{Overlaps}
\end{equation}
controls the asymptotic hybridization between oscillator wave functions 
(\ref{Radial-Wave-Function}) at positions ${\bf r}_i$ and ${\bf r}_j$ through 
the stiffness $s$. 

The model that has become the simplest paradigm for the synchronization phenomenon 
is the Kuramoto model \cite{Kuramoto1975,Kuramoto1987}. In terms of the Hamiltonian 
(\ref{Oscillator-Network}), the spectral distribution (\ref{Spectral-Density}), and the overlaps 
(\ref{Overlaps}), the Kuramoto model reads 
\begin{equation}
    \Dot{\theta}_i=\Omega_i+\sum_{j=1}^N K_{ij}\sin{(\theta_i-\theta_j)}+\zeta_i(t),
    \label{Kuramoto-Model}
\end{equation}
where $\theta_i$ is the phase of the $i-$th oscillator, and $\zeta_i$ are Gaussian 
noises, $\overline{\zeta_i(t)}=0$ and $\overline{\zeta_i(t)\zeta_j(t^\prime)}=2\gamma k_BT\delta_{ij}\delta(t-t^\prime)$,  
associated, via fluctuation-dissipation theorem, to the damping, ${\cal I}m[\Omega_i]\equiv\gamma\neq 0$,  
that controls the decay of the localized quadrupolar vibrations into thermal modes of the bath \cite{Sakaguchi1988}. 
For $N$ independent oscillators, $K_{ij}=0,\;\forall i,j$, the solution to Kuramoto's model in eq. (\ref{Kuramoto-Model}) 
is simply 
\begin{equation*}
\theta_i(t)=\Omega_i t + \phi_i,
\end{equation*}
where $\phi_i$ is an arbitrary phase constant. In this case, the second-quantized displacements at the 
independent quadrupolar oscillators are written in Heisenberg's representation as
\begin{equation}
\hat{\alpha}_{2,2c,i}(t)=\sqrt{\frac{\hbar}{2B\Omega_i}}(b^\dagger_{2,2c,i}e^{-i\theta_i(t)}+b_{2,\overline{2c},i}e^{i\theta_i(t)}).
\label{Secon-Quantization-Displacements-Network}
\end{equation}
Second order perturbation theory then gives rise to attractive, two-particle interactions 
at each independent oscillator, and to local superconductivity, as described earlier, being, 
however, incoherent and unable to produce bulk superconductivity.

On the other hand, when $K_{ij} \neq 0$, Kuramoto's order parameter, defined as \cite{Kuramoto1987}
\begin{equation}
    r \; e^{i\Theta(t)}=\frac{1}{N}\sum_{j=1}^N \; e^{i\theta_j(t)},
\end{equation}
quantifies the overall entanglement of the oscillator network, in terms of the relative strengths 
between fluctuations and hibridization. If the spread of $g(w)$ is larger than $K_{ij}$, which is the 
case of dilute, distant elastic insertions, $s^2|{\bf r}_i-{\bf r}_j|^2\gg 1$, then all oscillators perform 
individual cycles, all phases $\theta_i(t)$ are uniformly distributed over a circle $[0,2\pi]$, and $r=0$: 
synchronization is not possible. In Fig. \ref{Fig-Quadrupolar-Network} (left) we show a dilute, random 
collection of quadrupolar oscillators whose overlap is not enough to overcome the spread of their 
distributed, individual natural frequencies. As a result, although at each elastic inhomogeneity 
attractive two-particle interactions emerge, their local Cooper pairs do not exhibit phase coherence 
and bulk superconductivity is not possible. For stronger overlaps, $K_{ij}$, however, when elastic 
insertions are more abundant and closer together, $s^2|{\bf r}_i-{\bf r}_j|^2\sim 1$, larger number of 
oscillators start to group together into clusters of a definite phase, $\Theta$, and then $r\neq 0$: 
partial or full synchronization is achieved. In Fig. \ref{Fig-Quadrupolar-Network} (right) we show a 
larger, random collection of quadrupolar oscillators whose overlap is now enough to overcome the 
spread of the distributed, individual natural frequencies. In this case the attractive two-particle 
interactions at all elastic inhomogeneities give rise to coherent Cooper pairing and to bulk 
superconductivity. The surge of bulk superconductivity in random oscillator networks, 
as $N\rightarrow\infty$, can thus be described by Kuramoto's order parameter, $r$, which, by using 
a mean field approximation, $K_{ij}=K, \forall i,j$, is found as \cite{Sakaguchi1988}
\begin{eqnarray}
    r(\delta,T)&=&\sqrt{1-\frac{K_c(\delta,T)}{K}},\nonumber\\ 
    \frac{2}{K_c(\delta,T)}&=&
    \int_{-\infty}^\infty dw \,\frac{g(D(w)+\Omega)}{w^2+1},\nonumber
\end{eqnarray}
with $D(w)=\gamma k_BT w$ \cite{Sakaguchi1988}. Furthermore, the equivalence between Kuramoto's 
and the $XY$ model \cite{Sakaguchi1988,Uezu2015} allows us to identify $K$ as the superfluid 
density which controls a second order, synchronization phase transition for $K\geq K_c(\delta,T)$.

%
\begin{figure}[!t]
	\begin{center}
		\includegraphics[scale = 0.4]{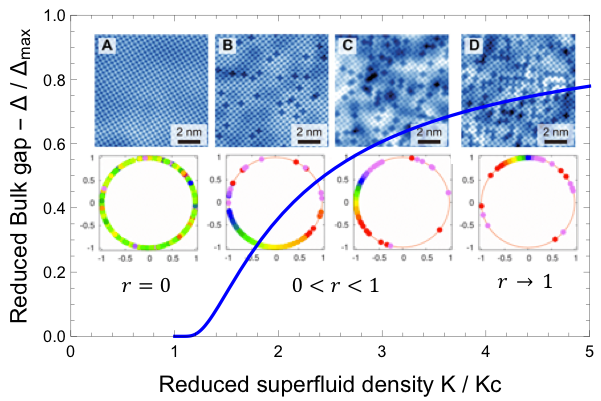}
	\end{center}
\caption{Bulk gap, $\Delta$, as a function of the superfluid density, $K$. 
(A) No inhomogeneities, $r=0$: no synchronization and $\Delta=0$. (B) 
and (C) Few inhomogeneities, $0<r<1$: partial synchronization and 
$\Delta<\Delta_{max}$. (D) Several inhomogeneities, $r\rightarrow 1$: 
full synchronization and $\Delta\rightarrow\Delta_{max}$. Insets from
refs. \cite{Hanaguri2018} and \cite{Mendoza2014}.}
\label{Fig-Synchronization}
\end{figure}
%

\section{Global Quadrupolar Superconductivity}

We can finally consider the problem of conduction electrons coupled to a globally 
connected, collective quadrupolar oscillator network. We start by writing down the
Hamiltonian
\begin{equation}
H=\sum_{{\bf k},\sigma}\xi_{\bf k}c^\dagger_{{\bf k},\sigma}c_{{\bf k},\sigma}+
b^\dagger_{2,2c}\cdot {\cal H}\cdot b_{2,2c}+H_{PVC},
\label{Global-Hamiltonian}
\end{equation}
where the PVC for our coupled oscillator network is now 
\begin{equation*}
H_{PVC}=\sum_{{\bf k},{\bf k}^\prime,\sigma}\sum_{ij}\Gamma^{ij}_{{\bf k},{\bf k}^\prime}
c^\dagger_{{\bf k}^\prime,\sigma}c_{{\bf k},\sigma}(b^\dagger_{2,2c,i}e^{-i\theta_j(t)}+ b_{2,\overline{2c},i}e^{i\theta_j(t)}).
\end{equation*}
This PVC Hamiltonian, which can be formally represented as $e^{i\hat{\theta}(t)}\cdot\Gamma\cdot\hat{\alpha}$,  
describes how the displacement, $\hat{\alpha}_i$, at the $i-$th insertion, is affected the phase, $e^{i\theta_j(t)}$, 
of the neighboring $j-$th insertion, through the matrix element
\begin{equation*}
    \Gamma^{ij}_{{\bf k},{\bf k}^\prime}=-\Gamma_0
    \int d^3{\bf r}\, \varphi_{{\bf k}^\prime}^*({\bf r}) \left[R_0\frac{\partial U(r_j)}{\partial r_i}Y_{2,2c}\right]
    \varphi_{{\bf k}}({\bf r}).
\end{equation*}
Upon synchronization the entire coupled network behaves as a single, entangled oscillator, whose 
spectrum, ${\bf \Omega}={\cal F}^\dagger\, {\cal H}{\cal F}$, is obtained by diagonalization of 
(\ref{Oscillator-Network}) or (\ref{Global-Hamiltonian}) through 
\begin{equation}
\hat{\alpha}^\prime={\cal F}^\dagger\hat{\alpha}, 
\label{Transformation}
\end{equation}
and possesses one lowest, stationary, $\sigma-$bonding eigenvalue, $\Omega_\sigma\in{\cal R}$,
that governs the asymptotic entanglement in the network \cite{Zambrini2013}. 
Transformation (\ref{Transformation}) applied to $e^{i\hat{\theta}(t)}\cdot\Gamma\cdot\hat{\alpha}$
produces, besides several, short-lived, transient couplings, an asymptotic coupling, $r\,e^{i\Theta(t)}\, \Gamma_\sigma^\prime \hat{\alpha}_\sigma^\prime$,
with $\Gamma^\prime={\cal F}^\dagger\cdot\Gamma\cdot{\cal F}$, 
that associates Kuramoto's order parameter to the lowest, stationary and asymptotic, $\sigma-$bonding 
eigenvector, $\hat{\alpha}_\sigma^\prime$. In Heisenberg's picture such {\it global} PVC can thus be
written as
\begin{equation*}
H_{PVC}=\sum_{{\bf k},{\bf k}^\prime,\sigma}r\, \Gamma^\prime_{{\bf k},{\bf k}^\prime}
c^\dagger_{{\bf k}^\prime,\sigma}c_{{\bf k},\sigma}(b^\dagger_{\sigma}e^{-i\Theta(t)}+ b_{\overline{\sigma}}e^{i\Theta(t)}),
\end{equation*}
where $b^\dagger_{\sigma},b_{\sigma}$ are bosonic, {\it global} phonon creation and anihilation operators
for the $\Theta-$phase-locked, globally entangled oscillator network, associated to the asymptotic, 
lowest frequency, $\Omega_\sigma$, corresponding to the $\sigma-$bonded eigenvector 
$\hat{\alpha}_\sigma^\prime$. 

The globally entangled form for the PVC produces, in second order perturbation theory and within 
the spirit of the adiabatic, Born-Oppenheimer approximation, an effective, real, attractive two-particle 
interaction which for electrons with $|\xi_{\bf k}|,|\xi_{{\bf k}^\prime}|\ll\hbar\Omega_\sigma$ is 
separable and can be written as
\begin{eqnarray}
\overline{V}_{{\bf k},{\bf k}^\prime}&=&-r^2 \; \overline{V}_0 \; \eta(\hat{\bf k})\eta(\hat{\bf k}^\prime) \;
\Theta(|\xi_{\bf k}|-\hbar\Omega_\sigma)\Theta(|\xi_{{\bf k}^\prime}|-\hbar\Omega_\sigma),\nonumber\\
\Delta_{\bf k}&=&\Delta\eta(\hat{\bf k})\, r\, e^{i\Theta},
\label{Convoluted-Interaction}
\end{eqnarray}
where $\overline{V}_0=|\Gamma_{{\bf k},{\bf k}^\prime}^\prime|^2/\hbar\Omega_\sigma$, 
corresponds to the convolution of all short range interactions at the local elastic insertions. 
From the self-consistent gap equation in eq. (\ref{Gap-Equation}) and the BCS interaction in eq. 
(\ref{Convoluted-Interaction}), the $T=0$ bulk gap thus becomes 
\begin{equation}
    \Delta =\frac{2\hbar\Omega_\sigma}{r(\delta,0) \sqrt{\left\langle \eta^2 \right\rangle_{FS}}} 
    \; \exp{\left\{-\frac{1}{r^2(\delta,0) \overline{\lambda} \langle\eta^2\rangle_{FS}}\right\}},
    \label{Bulk-Gap}
\end{equation}
where $\overline{\lambda}=N(\epsilon_F)\overline{V_0}$. The bulk gap, $\Delta$, is then a 
function of the superfluid density, $K$, through Kuramoto's order parameter, 
$r(\delta,0)=\sqrt{1-K_c(\delta,0)/K}$, and it's evolution across the synchronization 
phase transition is shown in Fig. \ref{Fig-Synchronization} for: (A) no inhomogeneities 
and $r=0$; (B)-(C) few inhomogeneities and $0<r<1$; and (D) several inhomogeneities 
and $r\rightarrow 1$ .

The critical temperature, in turn, is obtained as solution to the gap equation (\ref{Gap-Equation}) 
and the BCS interaction (\ref{Convoluted-Interaction}) at $\Delta=0$
\begin{equation}
    k_BT_c = 1.13\,\hbar\Omega_{\sigma}\, \exp{\left\{-\frac{1}{r^2(\delta,T_c)\overline{\lambda}\langle\eta^2\rangle_{FS}}\right\}},
    \label{eq:tc}
\end{equation}
which is a transcendental equation for $T_c$ given in terms of $r(\delta,T_c)=\sqrt{1-K_c(\delta,T_c)/K}$ . 
Nevertheless, the ratio between $\Delta$ and $T_c$ can be calculated for
$2\gamma k_B T_c\ll r^2(\delta,0)K$
\begin{equation}
    \frac{2\Delta}{k_BT_c} = \frac{3.53/\sqrt{\left\langle \eta^2 \right\rangle_{FS}}}{\sqrt{(1-K_c/K)}}
    \exp{\left\{\frac{2\gamma k_B T_c/K}{(1-K_c/K)^2\overline{\lambda}\langle\eta^2\rangle_{FS}}
    \right\}},
    \label{Bulk-Gap-to-T-ratio}
\end{equation}
and is exponentially larger than $2\Delta_{PG}/k_B T^*$ because of dissipation, 
$\gamma\neq 0$, associated to the decay of the collective quadrupolar vibrations 
into thermal modes of the bath
\begin{equation*}
2\Delta/k_B T_c\gg 2\Delta_{PG}w(k_F)/k_B T^*\;\mbox{for}\; r^2(\delta,T_c)\ll r^2(\delta,0).
\end{equation*}
%

%
\begin{figure}[!t]
	\begin{center}
		\includegraphics[scale = 0.4]{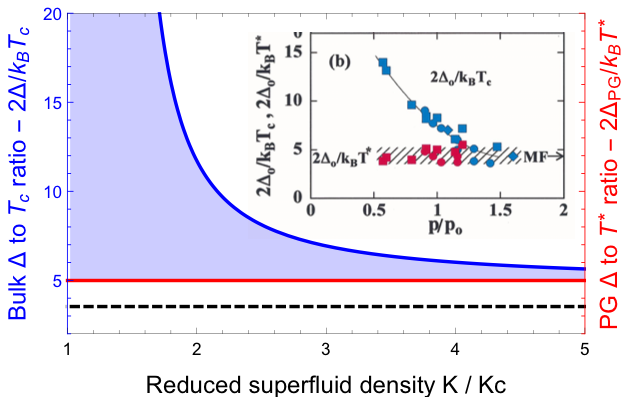}
	\end{center}
\caption{Bulk gap (top, left, blue) and pseudogap (center, right, red) ratios to $T_c$ 
and $T^*$, respectively, and BCS's universal ratio (bottom, dashed, black). Decay of
vibrations into bath modes results in $2\Delta/k_B T_c\gg 2\Delta_{PG}w(k_F)/k_B T^*$ 
(blue area). Inset from ref. \cite{Nakano1998}.}
\label{Fig-Universal-Ratios}
\end{figure}
%

\section{Comparison to Experiments}

In figure \ref{Fig-Universal-Ratios} we show the behaviour of (\ref{Pseudogap-to-T-ratio}) 
and (\ref{Bulk-Gap-to-T-ratio}) as functions of the superfluid density $K$. The 
$\Delta_{PG}$-to-$T^*$ ratio does not depend on $K$ and, apart from a nonuniversal 
factor, $w(k_F)$, it is given by $3.53/\sqrt{\langle\eta^2\rangle_{FS}}\approx 3.53\sqrt{2}=4.99$. 
The $\Delta$-to-$T_c$ ratio, in turn, depends exponentially on $K$. Even if in cuprates 
the Uemura plot suggests that the ratio $k_B T_c/K\approx const$ \cite{Uemura2002}, a 
$K$ dependence in (\ref{Bulk-Gap-to-T-ratio}) is provided by $r(\delta,0)=\sqrt{1-K_c/K}$
appearing both as prefactor and as $r^4(\delta,0)$ in the denominator of the exponential. For this reason the 
$\Delta$-to-$T_c$ ratio is considerably larger at low $K$ (within the underdoped regime) 
while it saturates at large $K$ (within the overdoped regime). Rough estimates for the 
maximum values for $T_c$, assuming full synchronization, $r\rightarrow 1$, are given in 
Table \ref{Tabela} for two layered, high-temperature superconducting cuprates. The fairly 
large values for $T_c$ obtained in our calculations can be traced back to the high normal 
frequencies of the collective quadrupolar vibrations, $\Omega_\sigma$, that correspond 
to the roots to eq. (\ref{Normal-Frequencies}), and lay far above Debye's. 

%
\begin{table}[!h]
        \begin{tabular}{|c||c|c|c|c|c|c|}\hline 
         Comp./Param. &  $\alpha$      & $\eta$    &  $\zeta$      & $R_0$(nm) & $\Omega_\sigma$ (eV) & $T_c$ (K) \\ 
         \hline\hline 
         YBa$_2$Cu$_3$O$_{7-x}$ & 0.50 & 0.50 & 0.624  & 0.4 &  0.056 & 86  \\ \hline
         Bi$_2$Sr$_2$CaCu$_2$O$_{8+x}$ & 0.54 & 0.54 & 0.609 & 0.3 & 0.049 & 74      \\ \hline
        \end{tabular}
         \caption{$\zeta$ values extracted from \cite{Lin1989}. Rough estimates for $T_c$ from (\ref{eq:tc}), 
         assuming full synchronization, $r\rightarrow 1$, nanometer-size insertions, and the weak-coupling parameter, $\overline{\lambda} = 0.8$. }
         \label{Tabela}
\end{table}
%

\section{Conclusion}

Localized vibrations in cuprates have been observed with Raman spectroscopy 
\cite{Litvinchuk1992}, inelastic neutron scattering \cite{Egami2007}, and ultrafast 
optical coherent spectroscopy \cite{Novelli2017}, but their role to the mechanism 
of superconductivity remains controversial. STM spectroscopy, on the other hand,
show unambiguously that local Cooper pairs form at elastic inhomogeneities seen
in STM topographic images, as shown in Fig. \ref{Fig-Inhomogeneities-BSCCO} \cite{Yazdani2007}. In this 
work, we showed that collective quadrupolar vibrations localized at random, elastic 
inhomogeneities, are able to capture several key features of unconventional superconductivity. 
Using concepts from elasticity theory, nuclear superfluidity, and the emergence of 
collective behaviour in oscillator networks, we have shown that the phenomena of 
phase fluctuations, pseudogap formation, local $d_{x^2-y^2}-$wave Cooper pairing,
and phase coherent, bulk high-temperature superconductivity arise naturally from a 
synchronization problem in nuclear oscillator networks. Furthermore, we especulate 
that the quadrupolar nature of the node oscillators may be associated to the nematicity 
observed in the pseudogap phase \cite{Daou2010,Fradkin2010,Sato2017}, while its 
double degeneracy, for weakly anisotropic systems, may be relevant to the understanding 
of the two-gap structure also observed in the pseudogap phase \cite{Yoshida2009}. 
We hope our formulation of unconventional superconductivity, may help to shed 
some light into the remarkable emergent phenomenon of unconventional 
high-temperature superconductivity.

\section{Acknowledgements}

MBSN acknowledges Roberta Zambrini for useful comments. This work was
supported by CAPES.

\end{document}